\documentclass[preprint]{aastex} 

 \usepackage[T1]{fontenc}
\usepackage{graphicx}        
\usepackage{amssymb}         
\usepackage[usenames,dvipsnames]{color}          
\usepackage{epstopdf}
\usepackage[normalem]{ulem}
\definecolor{darkblue}{rgb}{0,0,0.5}
\usepackage{bm}
\DeclareFontFamily{OT1}{pzc}{}
\DeclareFontShape{OT1}{pzc}{m}{it}%
             {<-> s * [1.1500] pzcmi7t}{}
\DeclareMathAlphabet{\mathscr}{OT1}{pzc}%
                                 {m}{it}

\DeclareGraphicsExtensions{.pdf,.png,.jpg,.tif,.pdf}

\usepackage{amsmath}

\renewcommand{\a}{{\mathbf{a}}}

\newcommand{\x}{\mathbf{x}}
\newcommand{\A}{{\mathbf{A}}}
\newcommand{\B}{{\mathbf{B}}}

\newcommand{\I}{{\mathbf{I}}}

\newcommand{\T}{{\mathbf{T}}}

  \renewcommand{\le}{\leqslant}

\allowdisplaybreaks[1]

\begin{document}


\title{The scattering of $f$- and $p$-modes from ensembles of thin magnetic flux tubes -- An analytical approach}

\author{Chris S.~Hanson}
\author{Paul S.~Cally}

\affil{Monash Centre for Astrophysics and School of Mathematical Sciences,\\ Monash University, Clayton, Victoria 3800, Australia}
  \email{christopher.hanson@monash.edu}

\shortauthors{C.S. Hanson \& P.S. Cally}

\begin{abstract}
Motivated by the observational results of \citet{braun_1995}, we extend the model of \citet{hanson_cally_2014} to address the effect of multiple scattering of $f$ and $p$-modes by an ensemble of thin vertical magnetic flux tubes in the surface layers of the Sun. As in observational Hankel analysis we measure the scatter and phase shift from an incident cylindrical wave in a coordinate system roughly centred in the core of the ensemble. It is demonstrated that, although thin flux tubes are unable to interact with high order fluting modes individually, they can indirectly absorb energy from these waves through the scatters of kink and sausage components. It is also shown how the distribution of absorption and phase shift across the azimuthal order $m$ depends strongly on the tube position, as well as on the individual tube characteristics. This is the first analytical study into an ensembles multiple scattering regime, that is embedded within a stratified atmosphere.
\end{abstract}



\section{Introduction}
The solar surface is threaded with thin magnetic filaments that often appear at granule or supergranule boundaries, or more prominently as massive ensembles (plage) associated with active regions. These filaments may be modelled as isolated thin flux tubes embedded within a field free plasma, interacting strongly with the solar acoustic $p$-modes and the surface gravity $f$-mode, both absorbing and scattering wave energy.  Their seismic signature is a powerful constraint on models of solar surface structure and magnetism. It is well established \citep[etc.]{braun_etal_1988,bogdan_etal_1993} that solar $f$- and $p$-modes interact strongly with sunspots, with up to $70\%$ of the incident power being absorbed and substantial phase shifts being observed in the remnant outgoing waves. Plage, which can be considered as an ensemble of distinct flux tubes, also absorbs, though more weakly  \citep[around 20\%;][]{braun_1995}. Although \citet{braun_1995} failed to identify any significant phase shift due to plage, later work \citep{chen_etal_1998,braun_birch_2008} found measurable time travel shifts of order 10 seconds, still considerably less then those observed within the penumbral and umbral regions (of order 20 and 40 seconds respectively).

While mechanisms such as mode-conversion \citep{cally_etal_2003} act in absorbing wave energy in larger magnetic features, it is believed that scattering regimes are the cause of the observed absorption within smaller fibril structures \citep{bogdan_fox_1991}.

Thus far, few analytical studies have investigated scattering from flux tube ensembles within gravitationally stratified atmospheres. Early models were restricted to non-stratified studies due to the mathematical complexity that gravity induces. Nevertheless, these models still demonstrated interesting characteristics of ensemble behaviour that could be used for determining subsurface structures. \citet{bogdan_zweibel_1985} and \citet{zweibel_bogdan_1986} showed that fibril ensembles can induce measurable frequency shifts. Furthermore \citet{bogdan_zweibel_1987} extended this work, showing that the multiple scattering of waves induces a cascade of acoustic power to ever smaller lengths scales as a wave propagates further into a fibril region. \citet{ryutova_1993,ryutova_1993_II} analysed wave propagation in unstratified models permeated by large numbers of non-identical flux tubes each having its own characteristic resonant frequency, finding 
significantly enhanced absorption. 

However, \citet{tirry_2000} argues that these resonances disappear when the flux tubes have a mechanism for downward energy loss, as in the stratified case. Within gravitationally stratified media, magnetic flux tubes exhibit both discrete and continuous spectra of oscillation, the former related to the $f$- and $p$-modes of the external medium, and the latter deriving from the acoustic jacket \citep{bogdan_cally_1995} of horizontally bound (`near field') slow waves that propagate vertically on the tube and take energy away. These play an important role in the multiple scattering regime \citep{hanasoge_2009}. Since stratification both requires the inclusion of the near-field modes, and possibly removes resonances, it then falls to the near-field multiple scattering regime to explain the observed enhanced scattering and absorption \citep{bogdan_fox_1991}. 

Recent years have seen several studies investigating scattering from single tubes \citep[etc]{hanasoge_etal_2008,hindman_jain_2012}. \citet{jain_etal_2009,jain_etal_2011b,jain_etal_2011} modelled ensembles within a stratified atmosphere, but neglected any scattering regimes and thus addressed the ensemble as a collective of individual non-interacting tubes. \citet{hanasoge_2009} were the first to outline the significance of the multiple scattering regime between pairs of tubes, by utilizing the analytical methods of \citet{kagemoto_yue_1986}, showing that the horizontally evanescent near-field plays a significant role in altering the scattered wave field coefficients when the tubes are in close proximity.  So far no analytical study has examined multiple scattering within an ensemble with gravity. However, a growing body of numerical studies have begun to shed light on the significance of this regime. \citet{felipe_2013} showed that the multiple scattering regime significantly impacts absorption coefficients, as well as generally decreasing the phase shift of the outgoing wave. It was also noted that the absorption generally increases with the number of tubes. \cite{daiffallah_2014} recently investigated the surface velocity profiles created by closely packed thin tube ensembles, showing that relative tube position and separation plays significant roles in the resultant vertical velocity profiles.

We \citep{hanson_cally_2014} applied a slightly modified version of the \citet{kagemoto_yue_1986} formalism to the interactions of $f$ modes with a pair of flux tubes ($p$-modes were included in the mathematical development, but were not investigated in detail because of their much weaker scattering). We restricted attention to the sausage and kink motions of the tube ($|m|\le 1$), as the mathematical thin tube approximation does not support higher order fluting modes.  In this paper we focus on the interactions between tubes that constitute a larger ensemble. In particular we are interested in the absorption and phase shift that can be determined from the outgoing cylindrical wave. Section~\ref{sec:maths} will outline the mathematical formalism important to this study, with Section~\ref{sec:results} presenting the results. Discussion and conclusions are given in Section~\ref{sec:concl}, with particular reference to the effect of the multiple scattering regime on measurable outgoing parameters. Higher order $m$ modes are included because they do interact with extended ensembles.
 
\section{Scattering Formalism}\label{sec:maths}
In this section we outline the scattering of various azimuthal modes  (characterized by integer $m$) from an ensemble of flux tubes. The formalism for the propagation of $f$-modes and their interaction with a pair of thin flux tubes is outlined in detail by \citet{hanson_cally_2014}. For the purpose of this study, we only mention the key parts of the formalism required to extend the model to an ensemble of arbitrarily positioned non-identical tubes. 

Consider a random assortment of vertical flux tubes embedded within a field free atmosphere. The atmosphere is a stratified adiabatic truncated polytrope (adiabatic index $\gamma=5/3$), with constant gravity. In such an atmosphere, the  complete acoustic wave field ($\Psi$) will consist of three components:
\begin{equation}\label{eq:wavefield}
\Psi=\Psi_{\textrm{inc}}+\Psi_{\textrm{sca}}+\Psi_{\textrm{int}},
\end{equation}
where inc, sca and int specify the incident, scattered and internal wave fields, respectively.
In the presence of magnetic filaments, an incoming $f$- or $p$-mode ($\Psi_{\textrm{inc}}$) will interact with each flux tube, and in turn be scattered back into the external medium ($\Psi_{\textrm{sca}}$). The fraction of energy that is scattered forms a wave field of radially propagating  modes, while the remainder of the incident wave energy is transported vertically along the tube axis ($\Psi_{\textrm{int}}$). Being a stratified atmosphere, $\Psi_{\textrm{sca}}$ describes both the scattered propagating and evanescent waves \citep{bogdan_cally_1995}. 

By adopting a right-handed cylindrical coordinate system, $\x=(r,\theta,s)$, we express the displacement eigenfunctions of the incident wave ($\Psi_{\textrm{inc}}$)  as
\begin{equation}
 \Psi_{\rm inc}(\x,t)=\sum\limits^{n_p}_{n=0}\sum\limits^{\infty}_{m=-\infty}i^mJ_m(k^p_nr)\Phi_n(\kappa_n^p;s)\,e^{i(m\theta-\omega t)},
\end{equation}
where $J_m(k^p_nr)$ is the Bessel function of order $m$, and $\Phi_n(\kappa^p_n;s)$ is the vertical displacement eigenfunction for a $n^{th}$ order $p$-mode with frequency $\omega$ and wavenumber $k_n$. The $f$-mode is characterized as a $n=0$ wave, with the $p_n$ mode spectrum truncated at $n=n_p$. A dimensionless depth parameter ($s=-z/z_0$) is used, which is scaled by the reference height of $z_0=-392$~km. 

Since we are only considering slender flux tubes, we utilize the thin tube approximation \citep{bogdan_etal_1996}, as the tube radius is small compared to the incident wavelength below $z_0$. This approximation reduces the number of oscillating modes within the tube to two, the sausage ($m=0$) and the kink ($|m|=1$). Assuming the tube is axisymmetric, incident waves of order $m$ will only excite and scatter into waves of the same $m$ order. The only mode scattering that occurs for isolated tubes is in $n$. Thus, the scattered wave field ($\phi^S_{\textrm{m}}$) from a thin tube will be
\begin{equation}\label{eq:scatter}
\phi^S_m(\x)=- \sum\limits_{n=0}^{n_p} S_{mn}\Phi_n(\kappa_n^p;s)H^{(1)}_m(k_n^pr)e^{im\theta} - \sum\limits_{n=n_p}^{N} S_{mn}\zeta_n(\kappa_n^j;s)K_m(k_n^jr)e^{im\theta} ,
\end{equation}
where $H^{(1)}_m(k^p_nr)$ and $K_m(k^j_nr)$ are the Hankel and K-Bessel functions of order $m$, $S_{mn}$ is the scattering coefficient and $\zeta_n(\kappa^j_n;s)$ is the vertical displacement eigenfunction of the jacket modes. We have assumed $e^{-i\omega t}$ time dependence. The exact method for matching the internal and external wave fields for the sausage and kink, and thus the determination of $S_{mn}$, is outlined in \citet{hanson_cally_2014} and references therein.

The scattering coefficient contains all the information needed to understand the wave interactions within an ensemble. Specifically, the absorption and phase shift can both be determined from $S_{mn}$. We define the absorption ($\alpha$) and phase shift ($\varphi$) of an ($m$,$n$) incident wave in the usual manner, 
\begin{equation}\label{eq:abs}
\alpha_{mn}=\frac{|A_{\textrm{in}}|^2-|A_{\textrm{out}}|^2}{|A_{\textrm{in}}|^2},
\end{equation}
\begin{equation}\label{eq:phase}
\varphi_{mn}=\textrm{arg}\left\{\frac{A_{\textrm{in}}}{A_{\textrm{out}}}\right\},
\end{equation}
where $A_{\textrm{in}}$ and $A_{\textrm{out}}$ are the complex amplitudes of the incoming and outgoing wave fields, respectively. In the case of energy outgoing into non-incident waves we use the energy fraction \citep{hindman_jain_2012},
\begin{equation}\label{eq:efrac}
\epsilon_{mn\rightarrow m'n'}=\left|\delta_{nn'}\delta_{mm'}+2S_{m'n'}\right|^2.
\end{equation}

In Hankel analysis \citep[etc]{braun_etal_1987,braun_etal_1988} the incoming and outgoing waves are determined by observing the waves within an annulus centered at a specific point. For comparative purposes we define the incoming and outgoing wave in Equations~(\ref{eq:abs}--\ref{eq:phase}) as waves centered at the coordinate origin.

In single tube cases the scatter is purely the product of the incident wave interacting with the tube. Within ensembles the interacting wave field will comprise of both the incident wave and the scattered waves from nearby tubes. Consequentially, the scattering coefficient of any flux tube is dependent upon all others, and vice versa. The simultaneous calculation of the coefficients requires the methodology of \citet{kagemoto_yue_1986}, and needs Equation~(\ref{eq:scatter}) to be expressed in matrix notation:
\begin{equation}\label{eq:onetube}
\phi_i^S=\sum\limits_n\left(A^T_i\Psi^S_i\right)_n,
\end{equation}
\begin{equation}
A_{in}=-\left (S_{-1n}~~S_{0n}~~S_{1n}\right ),
\end{equation}
\begin{equation}
(\Psi^S_{in})_{cd}=\begin{matrix}H^{(1)}_{c-2}(k^p_nr_i)\Phi_n(\kappa_n^p;s_d) & & (n\le n_p), \end{matrix}
\end{equation}
\begin{equation}
(\Psi^S_{in})_{cd}=\begin{matrix}K_{c-2}(k^p_nr_i)\zeta_n(\kappa_n^p;s_d) & & (n> n_p). \end{matrix}
\end{equation}
In this notation $c$ ranges over $[1,2|m_{max}|+1]$ where $m_{max}$ is the largest permitted $m$, $d$ over [1,250] and $s_d$ is the $d^{\rm th}$ point along the $s$ grid.

As Equation~(\ref{eq:onetube}) describes the contribution of the incident wave to the scatter ($\phi_0$), contributions to the coefficient from other tubes must also be considered. This is achieved firstly through relating a wave (incident or scattered) that is centered upon a point to a tube located at any other coordinate position. This requires the use of Graf's addition formula \citep{abram_1964} for the incident wave,
\begin{equation}\label{eq:incid}
J_m(k_n^pr_i)e^{im(\theta_i-\gamma_{il})}=
\sum\limits^{\infty}_{d=-\infty}J_{m+d}(k_n^pR_{il})J_d(k_n^pr_l)e^{id(\pi-\theta_l+\gamma_{il})},
\end{equation}
and the scattered waves,
\begin{equation}\label{eq:grafh1}
H^{(1)}_m(k_n^pr_i)e^{im(\theta_i-\gamma_{il})}=
\sum\limits^{\infty}_{d=-\infty}H_{m+d}^{(1)}(k_n^pR_{il})J_d(k_n^pr_l)e^{id(\pi-\theta_l+\gamma_{il})},
\end{equation}
\begin{equation}\label{eq:grafk}
K_m(k_n^jr_i)e^{im(\theta_i-\gamma_{il})}=
\sum\limits^{\infty}_{d=-\infty}K_{m+d}(k_n^jR_{il})I_d(k_n^jr_l)e^{id(\pi-\theta_l+\gamma_{il})}.
\end{equation}
In these expressions $R_{il}$ is the distance between tube centers, while $\gamma_{il}$ is the angular distance from the positive $x$ axis to the line of separation (See Figure~\ref{fig:coordhex}). Close examination of Equations~(\ref{eq:incid}--\ref{eq:grafk}) shows that any tube located at the coordinate origin of a $m$ incident wave ($R_{il}=0$) will experience a pure $m$ wave, while any tube that is not located at the wave's coordinate origin ($R_{il}>0$ ) will experience the wave as a mixture of all $m$ components. The same principle applies for the scattered propagating and evanescent waves (Equations~\ref{eq:grafh1}--\ref{eq:grafk}).

The scattered wave field from tube $i$ is then related as an incident wave on tube $l$ through the use of the \emph{Transformation Matrix} $\T_{il}$,
\begin{equation}\label{eq:TPSI}
\Psi^S_{in}=\T^n_{il}\Psi_{ln}^I,
\end{equation}
where the superscripts $S$ and $I$ specify a scattered and incident wave field, respectively. The elements that populate $\T_{il}$ are determined from Equations~(\ref{eq:grafh1}--\ref{eq:grafk}) and are detailed in \citet{hanson_cally_2014}. Taking the incident and nearby scattered wave fields into consideration, the complete scattered wave field from tube $l$ is a combination of Equations~(\ref{eq:onetube}) and (\ref{eq:TPSI}),
\begin{equation}\label{eq:20}
\phi_l=\sum\limits_n\left(a^T_{ln}+\sum\limits_{i=1,i\ne l}^{N}A_{in}^T\mathbf{T}_{il}^n\right)\Psi_{ln}^I,
\end{equation}
where $a^T_{ln}$ is a vector populated with the amplitudes of the incident wave (Equation~\ref{eq:incid}), and the internal summation is the contribution of all other scatterers.

\begin{figure}
\centering
\includegraphics[scale=0.4]{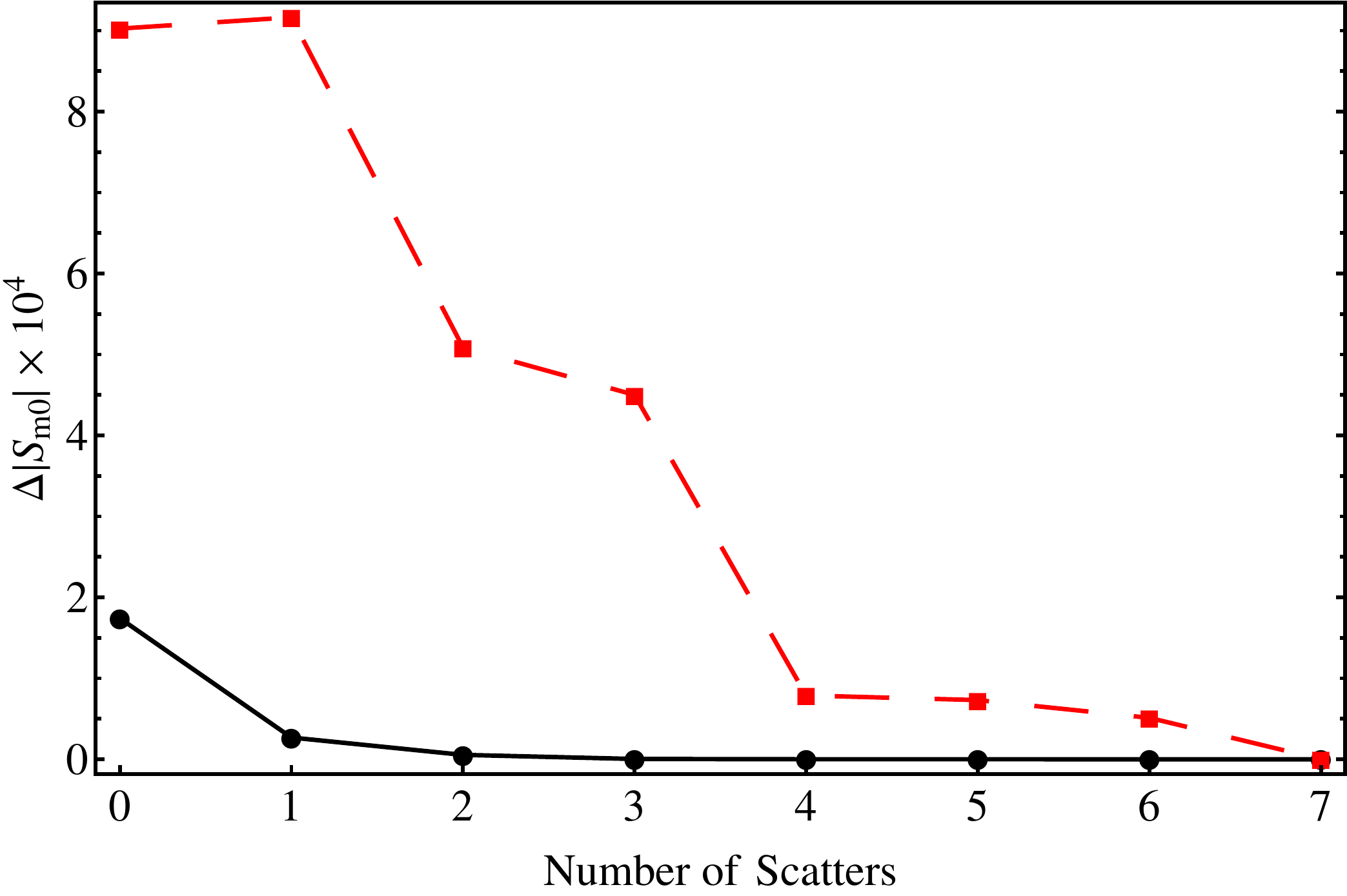}
\caption{The change in the scattering coefficient as the permitted number of scatters between tubes is increased. The $m=0$ (black) scatter coefficient converges quickly within 2--3 scatters, while the $|m|=1$) wave (red dashed) is not as quick to converge.}
\label{fig:noscatter}
\end{figure}

In short, and while referring to \citet{kagemoto_yue_1986} and \cite{ hanson_cally_2014} for details, the scattering coefficients for tube $l$ must be related to the scattering coefficients of an isolated case. Specifically, there exists a \emph{Diffraction Transfer Matrix} $\B_l$ such that the following holds true:
\begin{equation}
\A_l=\B_l\phi_l,
\end{equation}
where $\A_l$ is the vector containing the scattering coefficients for all $m$ and $n$ modes. From this Equation~(\ref{eq:20}) is rewritten,
\begin{equation}\label{eq:scatmatr}
\A_l=\B_l\left(\a_{l}+\sum\limits_{i=1,i\ne l}^{N}\mathbf{T}_{il}^T\A_i\right).
\end{equation}
In the case of two tubes \citep{hanson_cally_2014,hanasoge_2009} $\A_l$ can be easy calculated using a linear solve algorithm by rearranging Equation~(\ref{eq:scatmatr}). We extend the interaction between tube pairs to multiple $N$ tubes, namely:
\begin{equation}\label{eq:ntubeslinear}
\left[\I-\sum\limits_{i\ne l,l=1}^{N}\B_l\T^T_{il}\B_i\T^T_{li}\right]\A_l=\B_l\left(\a_l+\sum\limits_{i\ne l,l=1}^{N}\T^T_{il}\B_i\a_i\right).
\end{equation}
In this way we have calculated the exact $\A$ solution for $N$ tubes. Recent numerical studies (e.g. \citet{felipe_2013}) restrict the number of scatters between tubes to best mimic a complete scattered wave field, within the ensemble. Figure~\ref{fig:noscatter} shows the change in the scattering coefficient from the exact solution as the number of scatters increases. The $m=0$ scattering coefficient converges rapidly within three scatters, whilst the $|m|=1$ mode converges less quickly within seven scatters.

\subsection{The interaction of higher order $m$ modes within ensembles}
The Hankel analysis of plage \citep{braun_1995} has shown measurable non-zero absorption coefficients for $|m|>1$ waves. Due to the implementation of the thin tube approximation, we only permit the interaction of $|m|\le1$ waves with the tubes. Any higher order $m$ incident waves will not interact with the individual tubes, resulting in the scattering coefficient for that tube vanishing. Relating to Equation~(\ref{eq:wavefield}), the total wave field after a fluting mode ($|m|>1$) interacts with a thin tube located at the origin, will have a zero $\Psi_{\textrm{sca}}$ and $\Psi_{\textrm{int}}$ terms,
\begin{equation}\label{eq:wavefieldmg1}
\Psi=\Psi_{\textrm{inc}}.
\end{equation}
Consequently, the incident wave leaves the system unaffected by the thin tube. However, Graf's addition formula (Equations~\ref{eq:incid}--\ref{eq:grafk}) defines an $m$ mode by the coordinate origin. Any fluting mode defined with centre at the origin is experienced as a mixture of all other $m$ modes by any tube that is not located at the origin. Thus, as a fraction of the incident wave's energy is scattered by off-center tubes through $|m|\le 1$ interactions, the amplitude of the outgoing wave will be altered, despite the tubes being unable to interact with pure $|m|>1$ modes. In terms of an energy budget, the scattering from an off-center tube of $|m|\le1$ components reduces the outgoing power of the fluting modes.

In regards to the formalism of \citet{kagemoto_yue_1986}, the $\T_{il}$ and $\a_l$ are expanded to be fully populated for a larger $|m|$ set of modes. The interaction of the tubes with fluting modes depends solely on the terms in the $\B$ matrix. Since the tubes do not scatter these higher order modes, the coefficients that populate $\B$ and correspond to $|m|>1$ will simply be zero. In doing this there will be no scatter from an isolated tube located at the origin for a $|m|>1$ wave,
\begin{equation}
\A_l=\B_l\a_l=\mathbf{0}.
\end{equation}
 However, as $\T_{il}$ is fully populated (and $\a_{l}$ for off-center tubes) energy transfer from a $|m|>1$ wave to the tube is possible through kink and sausage components seen by off center tubes. In a closely packed ensemble any scatter from the tube at the origin will purely be due to neighboring tubes:
\begin{equation}
\A_l=\sum\limits_{i=1,i\ne l}^{N}\B_l\mathbf{T}_{il}^T\A_i.
\end{equation}


\section{Results}\label{sec:results}
\subsection{Symmetry Studies}
Let us consider ensembles of identical flux tubes, all with uniform plasma-$\beta=1$, that are symmetrically placed around the coordinate origin. In all these cases we have allowed Graf's formula to range over $|m|\le4$ to demonstrate the non-zero scattering of fluting modes. The incident wave is a $f$-mode of order $m$, with a wavelength of $\lambda=4.9$~Mm  and frequency $3$~mHz. A large discrete subset of jacket modes is chosen to mimic the continuum of modes present in an infinitely deep atmosphere. We define the change in absorption ($\Delta\alpha$) and in phase ($\Delta\varphi$) to be the difference between the values obtained when the tubes are interacting, and when they are not.

We begin by extending the model of \citet{hanson_cally_2014} to three tubes, aligned along the $x$ axis and centered upon the origin. In this ensemble we have two cases. Firstly, the tubes are positioned at $x=0$ and $\pm0.2\lambda$, which is close enough for near field interaction. In the second case,  $1.5\lambda$ is the separation distance between them. At this distance, the tubes are far enough apart to only interact through the far-field. The resultant absorption and phase shift for both cases are seen in Figure~\ref{fig:tir}. In each case, the absorption coefficient and phase shift peaks for $|m|\le 1$ incident waves, with the peaks being greatest for the near-field interaction case. If we compare both cases to their respective non-interacting coefficients\footnote{ Here we mathematically allow only a single scatter off each tube by neglecting the $\mathbf{T}_{il}^n$ terms in Equation (\ref{eq:20}).}, the changes in the absorption and phase shift become negligible when only interacting through the far-field. The changes in absorption for the near-field interacting triplet are of the order $10^{-2}$ compared to that of the isolated cases. However, when interacting through the far-field, the changes are of the order $10^{-3}$. This difference between the near- and far-field cases can also be seen in the phase changes. We conclude for this first ensemble set that for $|m|>1$ waves the scattering is non-zero despite the tubes not interacting directly with these waves. But, these scattering effects rapidly diminish as $|m|$ increases.
\begin{figure}
\centering
\includegraphics[scale=0.5]{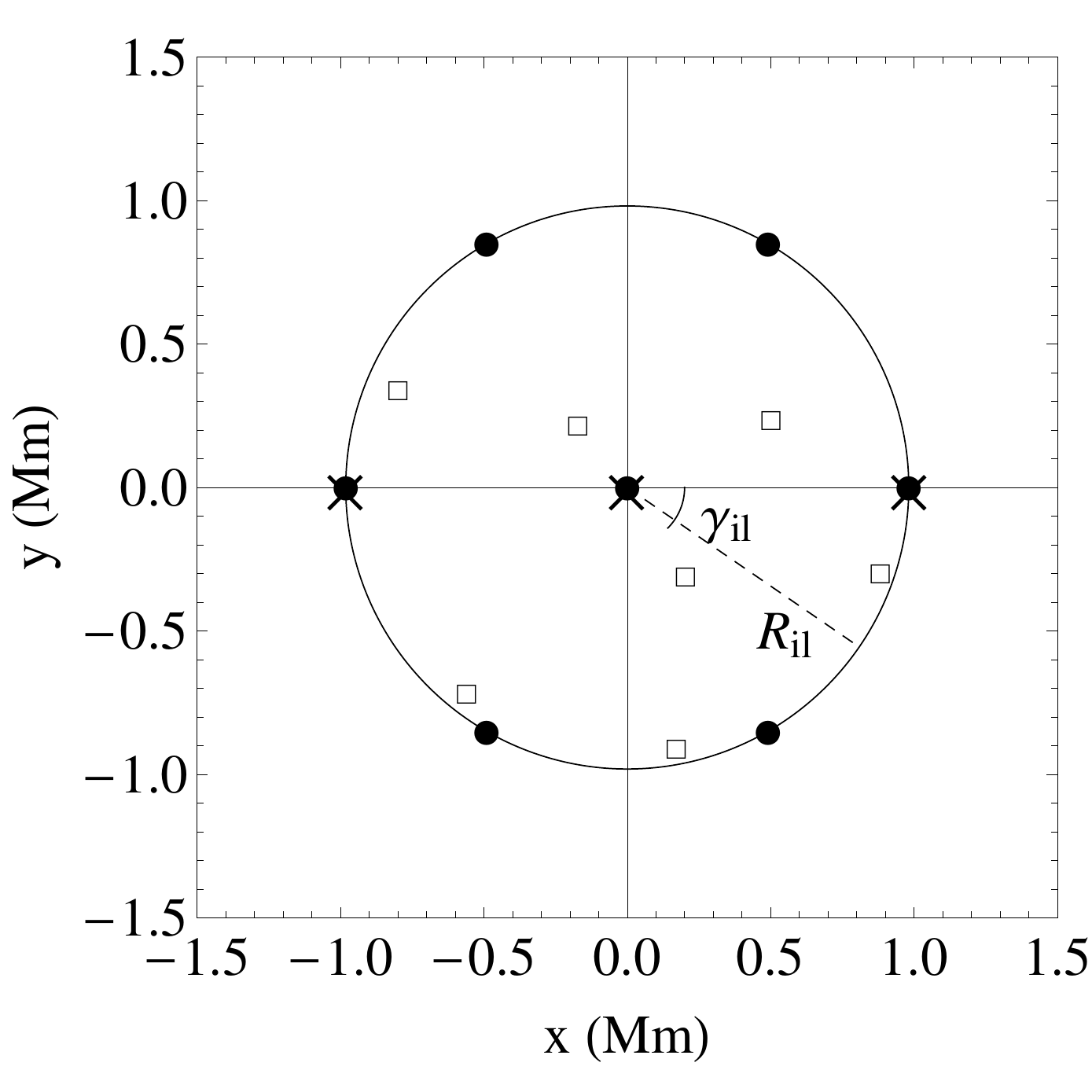}
\caption{The tube positions for all ensembles in this study. The dot points represent the positions of the 7 tube ensemble, as well as the six tube ensemble with the tube located at (0,0) absent. The crosses specify the positions of the three tubes ensemble (that are separated by $0.2\lambda$), while the squares indicate the positions of the tubes within the randomly positioned ensembles. The separation parameters $R_{il}$ and $\gamma_{il}$ are also represented. The large circle highlights the $1$~Mm radius circle inside which the random tubes cases are positioned.}
\label{fig:coordhex}
\end{figure}

\begin{figure*}
\centering
\includegraphics[width=\textwidth]{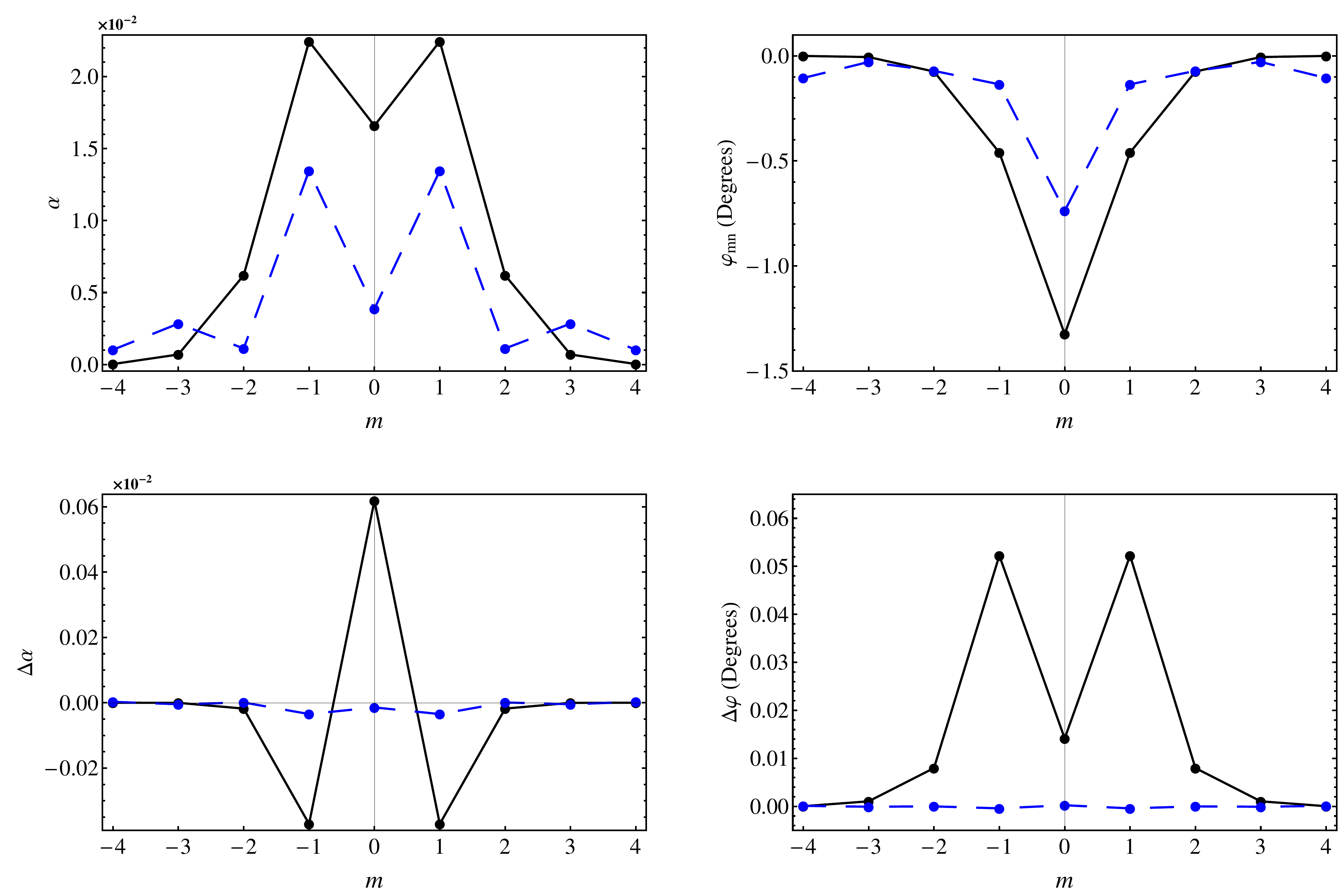}
\caption{The absorption and phase shift of the outgoing $f$-mode for three tubes aligned along the $x$ axis, centered on the origin and separated by 0.2~$\lambda$ (Black) and 1.5~$\lambda$ (Blue Dashed), where $\lambda=4.9$~Mm. The top panels show the absorption and phase shift of an incident $m$ wave when the three tubes are interacting. The bottom panels show the difference in scattering between interacting and non-interacting tubes. The change in absorption and phase is large when the tubes are interacting through the near field, but when far apart the interaction between the tubes is insignificant to the resultant outgoing wave.}
\label{fig:tir}
\end{figure*}

\begin{figure*}
\centering
\includegraphics[width=\textwidth]{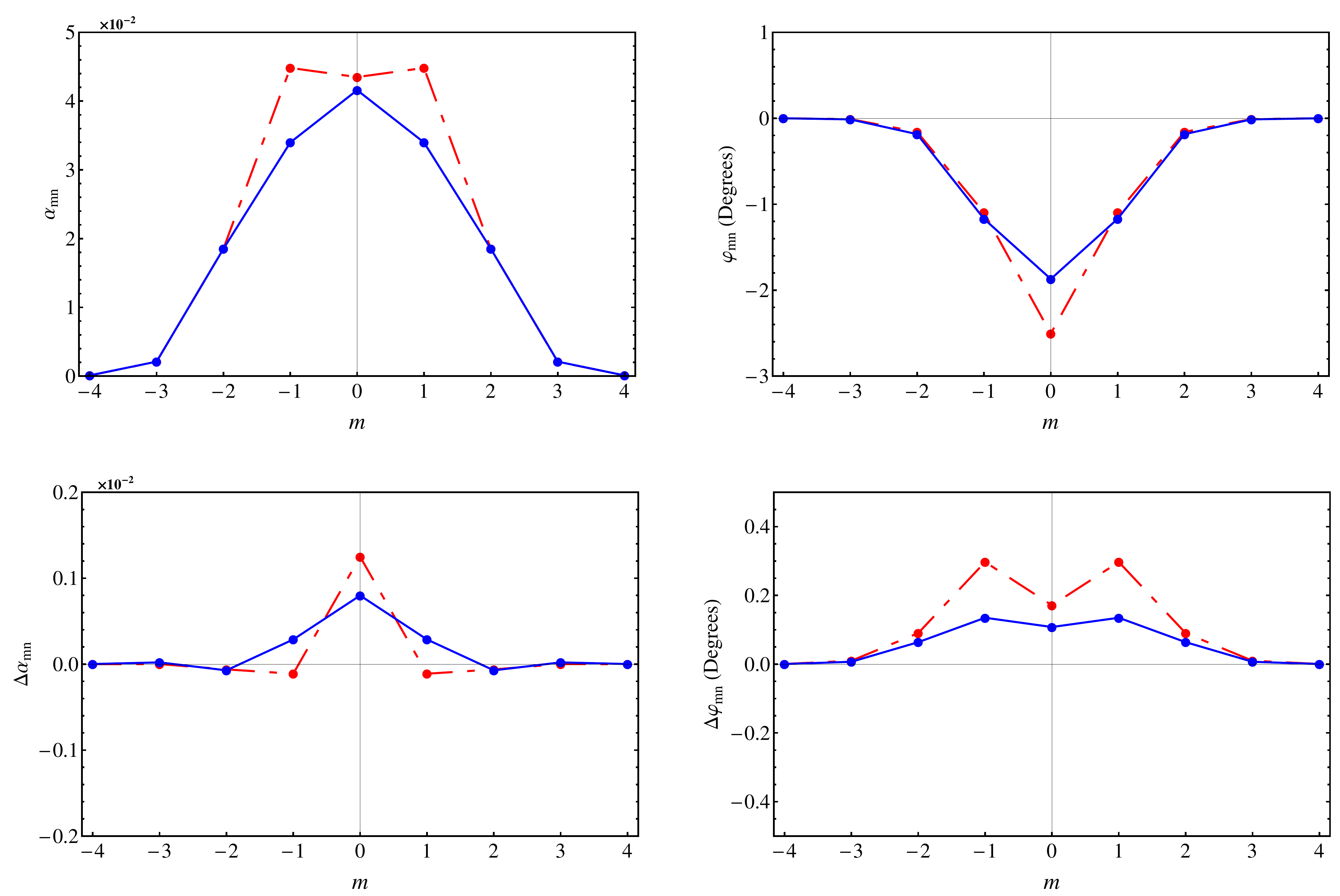}
\caption{The absorption (left) and phase of the outgoing wave (right) from a hexagonal ensemble of seven tubes (red dashed) and six tubes (blue). The scattering coefficient examined here is the $f$-$f$ coefficient. Six tubes are $0.2\lambda$ from the origin, with the seventh tube located at the center. The hexagonal distribution is symmetric and hence has symmetry between $\pm m$ modes. The addition of the seventh tube acts to increase absorption and shift the phase further to the negative, while also reducing $\Delta\alpha$ for the $m=\pm1$ modes. 
}
\label{fig:hex}
\end{figure*}

Extending the model to a larger ensemble, we also investigate the case of  many tubes positioned evenly around a circle of radius $0.2\lambda$. In particular we examine two similar ensembles, see Figure~\ref{fig:coordhex}. Firstly, six tubes are positioned on the circle, with the tube centers mapping out the vertices of a regular hexagon. In the second ensemble, a seventh tube is also positioned at the coordinate origin, as in \citet{daiffallah_2014}. The affects of the seventh tube within the circle of tubes is seen in Figure~\ref{fig:hex}. In each ensemble the absorption and phase shifts have slightly differing behavior for all $m$. The seventh central tube acts to increase the absorption of the ensemble for $|m|\le1$, but only enhances the phase shift of the $m=0$ wave. However, in comparison to their respective non-interacting values, the six tube case sees greatest change in absorption for  $|m|\le1$ waves, while the seven tube ensemble only peaks at $m=0$. The change in phase shift is significant for $|m|\le3$ for both ensembles, with the seven tube case experiencing a greater change. Comparison to the three tube ensemble shows that the absorption and phase shift are of the same magnitude. However, comparing the change in scattering properties from the non-interacting tube cases demonstrates that the larger ensembles experience a change in both phase and absorption that is an order of magnitude larger than the smaller three tube ensemble.

\begin{figure*}
\centering
\includegraphics[width=\textwidth]{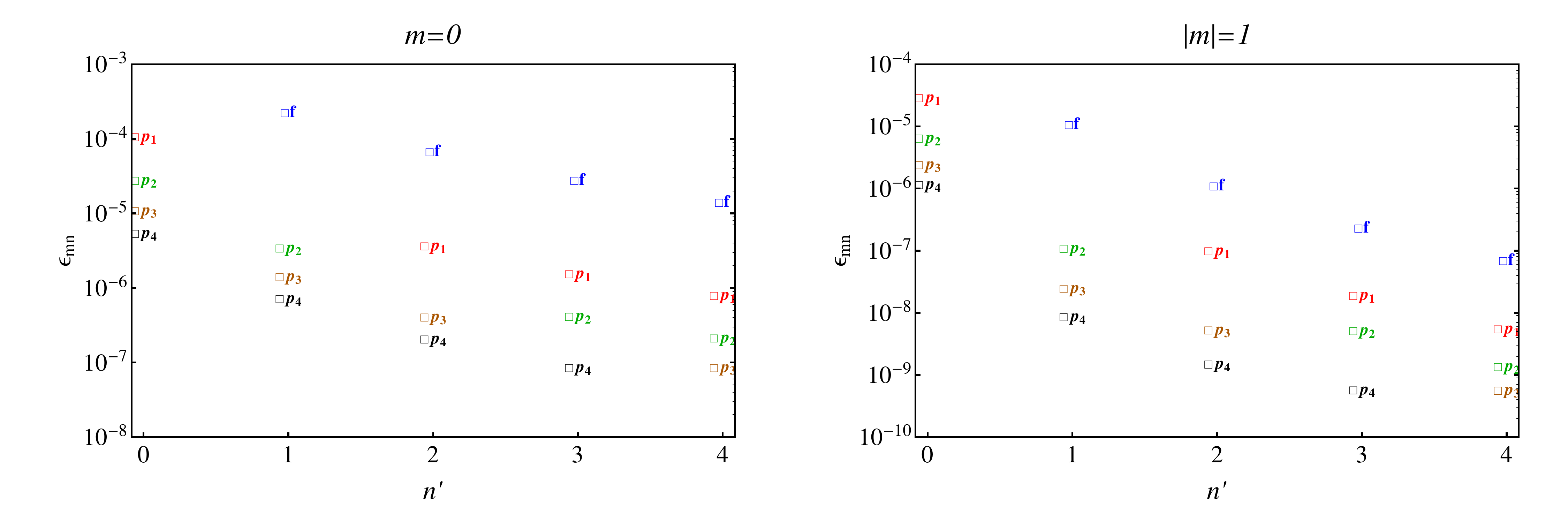}
\caption{The energy fraction (Equation~\ref{eq:efrac}) of the outgoing $p_{n'}$ from the seven tube hexagonal case, with an incident $p_n$ mode of frequency 3~mHz. The scattering is strongest when involving the $f$ mode, with the energy rapidly diminishing with increasing $n$ for both the sausage and kink modes. The color and small label indicate the nature of the incident wave ($f$, $p_1$, \ldots), the horizontal position indicates the scattered wave (for the incident $m$ only), and the vertical position represents the fractional energy in the scattered mode (logarithmic scale). The outgoing energy fraction in the original mode is not shown as it is nearly 1 in all cases.}
\label{fig:ppscat}
\end{figure*}

Within the solar atmosphere, as well as in this model, the scattering between tubes is not just restricted to the order ($m,n$)  of the incident wave. In fact the tubes may scatter into all other $m'$ and $n'$ wave components. The fraction of outgoing energy that is transferred from a $p_n$ to a $p_{n'}$ wave, for the above mentioned seven tube ensemble, is seen in Figure~\ref{fig:ppscat}. The plot is similar to Figure 3 of \citet{hindman_jain_2012} (for the $m=0$ mode),  demonstrating that for both the sausage and kink modes the energy fraction diminishes with increasing $n'$. The greatest energy transfer occurs with waves that are scattering from or to an $f$ mode. Furthermore,  Table~\ref{table:mmscat} shows the energy fraction of the outgoing wave components ($m'$) from an incident positive $m$. The strongest scattering is into the incident mode ($m'=m$), which is due to the $\delta_{mm'}$ in Equation~(\ref{eq:efrac}). Interestingly, due to the symmetry of the system, the scattering of an even $m$ is restricted to even $m'$ components, and vice versa for odd $m$.

\begin{deluxetable}{cccccc}
\tablecaption{The energy fraction $\epsilon_{mn}$ of outgoing wave components $m'$\label{table:mmscat}}
\tablecolumns{4}
\tablewidth{0pt}
\tablehead{$m'$&$m=0$&$m=1$&$m=2$&$m=3$&$m=4$}
\startdata

$-4$ & $4.3 \times 10^{-9}$ & $-$ & $7.8 \times 10^{-9}$ & $-$ & $1.5 \times 10^{-13}$\\ 

$-3$ & $-$ & $1.6 \times 10^{-7}$ & $-$ & $1.3 \times 10^{-8}$ & $-$\\ 

$-2$ & $2.0 \times 10^{-7}$ & $-$ & $5.7 \times 10^{-6}$ & $-$ & $7.8 \times 10^{-9}$\\ 

$-1$ & $-$ & $1.6 \times 10^{-6}$ & $-$ & $1.4 \times 10^{-7}$ & $-$\\ 

$0$ & $\mathbf{9.5 \times 10^{-1}}$ & $-$ & $3.5 \times 10^{-8}$ & $-$ & $5.4 \times 10^{-9}$\\ 

$1$ & $-$ & $\mathbf{9.5 \times 10^{-1}}$ & $-$ & $7.0 \times 10^{-10}$ & $-$\\ 

$2$ & $2.0 \times 10^{-7}$ & $-$ & $\mathbf{9.8 \times 10^{-1}}$ & $-$ & $1.8 \times 10^{-11}$\\ 

$3$ & $-$ & $1.2 \times 10^{-8}$ & $-$ & $\mathbf{9.9 \times 10^{-1}}$ & $-$\\ 

$4$ & $4.3 \times 10^{-9}$ & $-$ & $1.2 \times 10^{-10}$ & $-$ & $\mathbf{9.9 \times 10^{-1}}$
\enddata
\tablecomments{The energy fraction (Equation~\ref{eq:efrac}) of an outgoing $m'$ wave, produced by an incident $+m$ wave. The dominant diagonal (bold face) appears for cases where $m'=m$. The symmetry of the seven tube system forces the system to scatter into even $m'$, for an incident even $m$ wave. This is also the case for odd $m$. If the incident wave was of order $-m$ the results would be reversed about $m=0$.}
\end{deluxetable}

\subsection{Random Ensembles}
While symmetry studies are useful in determining some characteristics of ensembles that affect absorption and phase, we are also interested in how the random nature of an ensemble affects measurable parameters. We have built three ensembles, from which we will examine the effect of random positions, as well as various tube characteristics. The first ensemble is of seven identical $\beta=1$ tubes, positioned randomly within 1~Mm of the origin. The second and third ensembles are identical in tube placement to the first, but have a selection of different $\beta$. 

\begin{figure*}
\centering
\includegraphics[width=\textwidth]{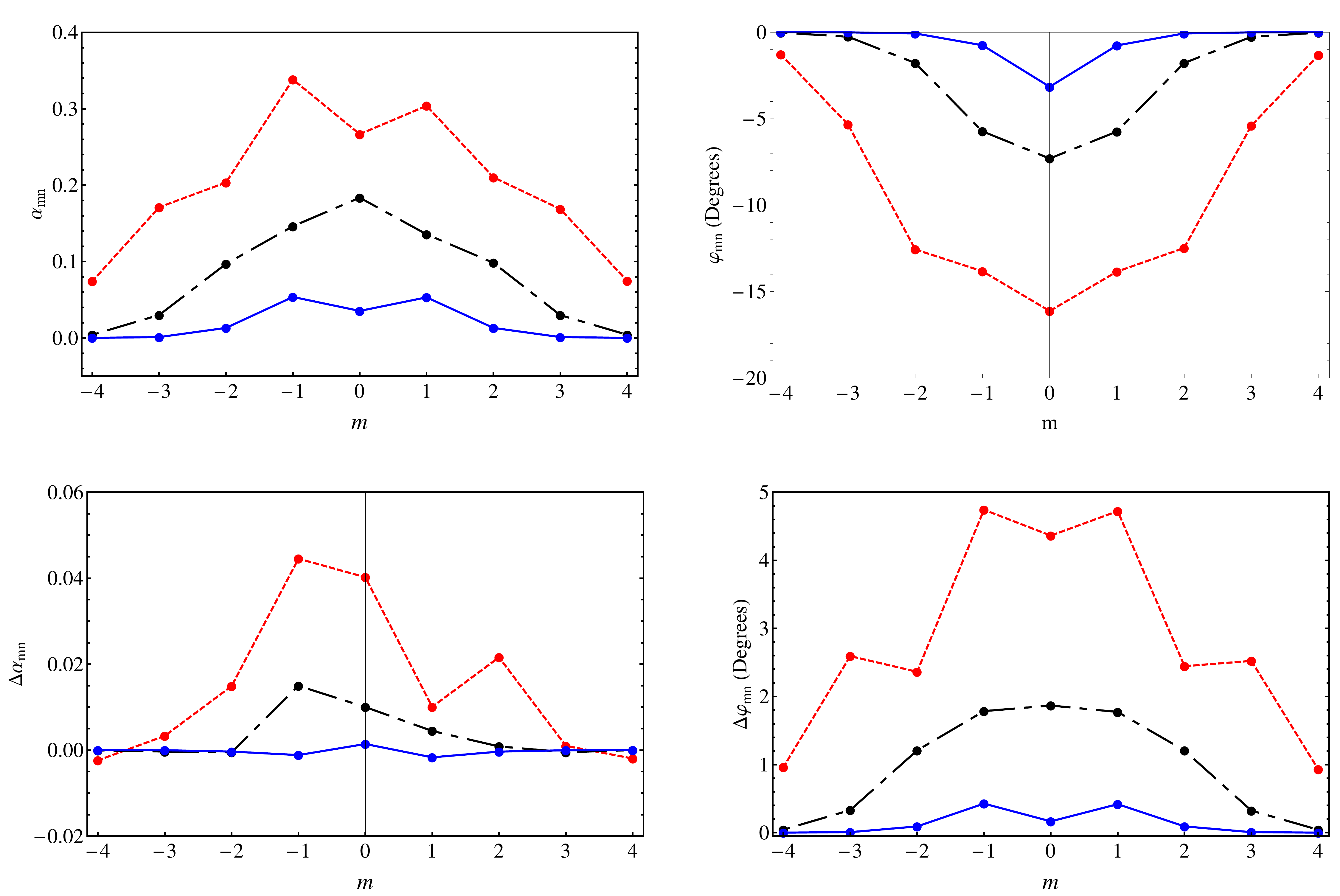}
\caption{The absorption and phase from an ensemble of seven randomly positioned identical ($\beta=1$) tubes. Again the we investigate $f$-$f$ mode scattering. The incident wave (blue solid: 3~mHz, black dot-dashed: 4~mHz, red dashed: 5~mHz) is scattered differently by the ensemble for different frequencies. The degree of absorption and phase shift is generally increased by higher frequency waves. The absorption of the incident wave is not symmetric for $\pm m$ modes, as the assortment is randomly positioned around the coordinate origin. However, the phase shift appears to be symmetrically distributed across $m$.}
\label{fig:random3freq}
\end{figure*}

In the first ensemble (randomly distributed identical tubes), three cases are studied for the incident wave frequencies of 3, 4 and 5~mHz. The resultant absorption and phase shift can be seen in Figure~\ref{fig:random3freq}. The largest scattering effects in both outgoing values and change from isolated values are seen in higher frequencies. The higher frequency waves also act in highlighting differences in the absorption coefficients between $\pm m$. As the tubes are not symmetrical around the origin a difference in absorption appears between $\pm m$. Interestingly, the $\pm m$  modes have no apparent difference in the phase shift. Similarly to the previous ensembles, the scattering coefficients peak at $|m|\le 1$ and rapidly decrease with increasing $|m|$. 

\begin{figure*}
\centering
\includegraphics[width=\textwidth]{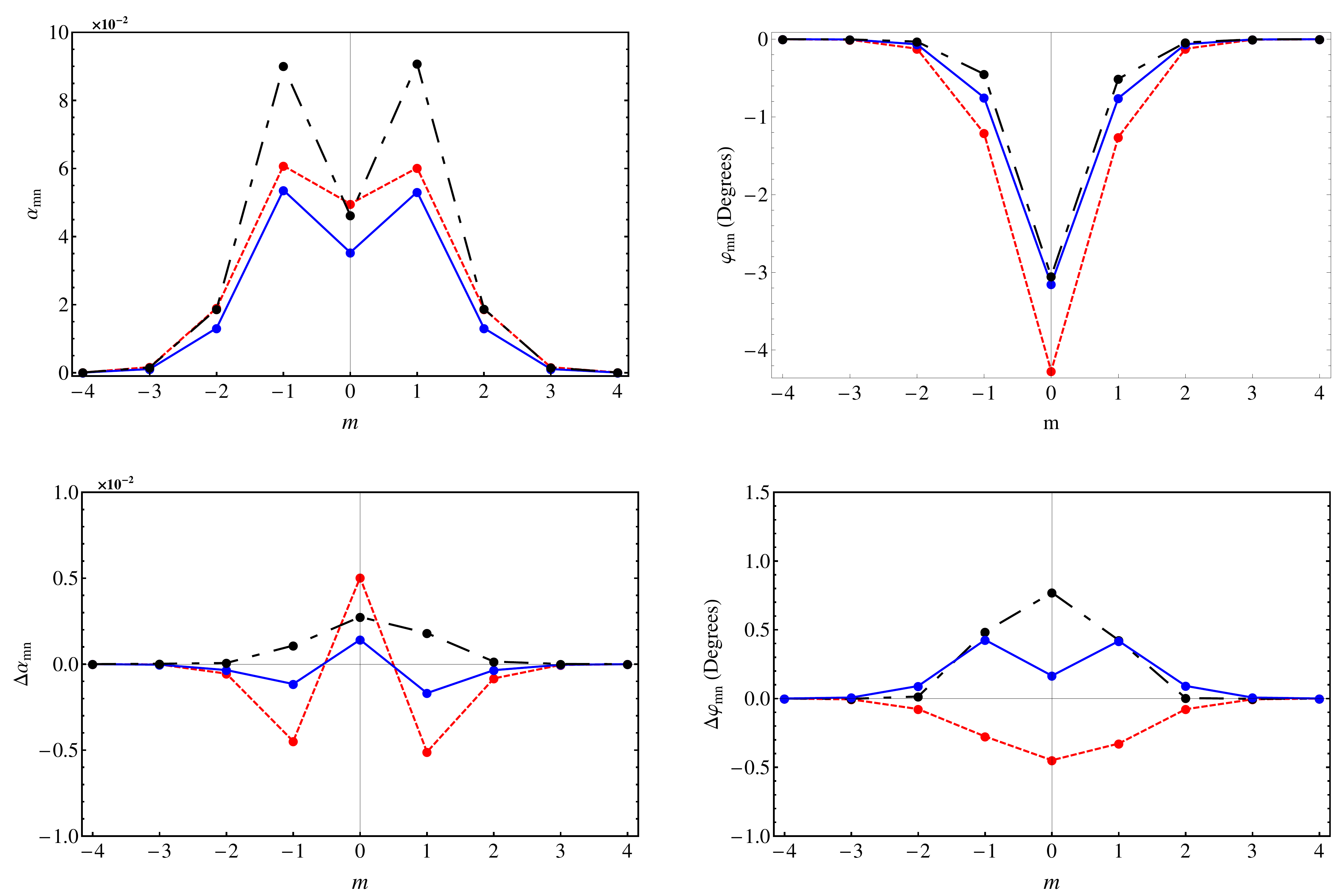}
\caption{The absorption and phase from an ensemble of identical tubes (blue) and two random non-identical ensembles (red dashed and black dot-dashed). The tube positions are identical to those of Figure~\ref{fig:random3freq}, but vary in $\beta$. The incident wave is a $f$-mode of 3~mHz and all of the tubes reside within 1~Mm of the origin. The absorption and phase are distributed differently across $m$ and are strongly dependent upon the individual tubes present. Differences in phase for $\pm m$ are more apparent when the tubes are non-identical.}
\label{fig:randomrandom}
\end{figure*}

The scattering effects of the second and third ensembles of non-identical flux tubes is shown in  Figure~\ref{fig:randomrandom}. Two cases are explored here, both with different selections of plasma $\beta$ (see Table~\ref{table:betatab}), and we compare these values to those of the ensemble of identical tubes. In both these cases the differing $\beta$ amongst the ensemble acts in enhancing the overall absorption when compared to the identical ensemble. However, the changes from the non-interacting ensemble demonstrates that depending on the $\beta$ of the tubes present, the distribution of $\Delta\alpha$ can be very different across $m$. In regards to the phase, one ensemble shows increased negative phase shift, while the other shows a slightly diminished phase shift from the identical ensemble case. The change in phase from isolated values supports this, showing that the multiple scattering between the tubes enhances this difference. We note here that while higher frequencies could not reveal the difference in phase across $\pm m$, the random distribution in $\beta$ highlights that a difference does exist (even if it is very small). 

\begin{deluxetable}{rrccc}
\tablecaption{The tube position and corresponding $\beta$ for the three random ensembles of seven fluxtubes.}
\tablecolumns{4}
\tablewidth{0pt}
\tablehead{$x$ (Mm)&$y$ (Mm)&$\beta$ (Case 1)&$\beta$ (Case 2)&$\beta$ (Case 3)}
\startdata
0.17 & -0.90 & 1 & 1 & 10 \\
0.20 & -0.30 & 1 & 5 & 0.5 \\
0.50 & 0.25 & 1 & 0.5 & 5 \\
-0.17 & 0.23 & 1 & 1 & 0.1 \\
-0.56 & -0.70 & 1 & 0.1 & 1 \\
-0.80 & 0.35 & 1 & 1 & 0.1 \\
0.82 & -0.29 & 1 & 0.1 & 0.5 
\enddata
\label{table:betatab}
\end{deluxetable}

\begin{figure*}
\centering
\includegraphics[width=\textwidth]{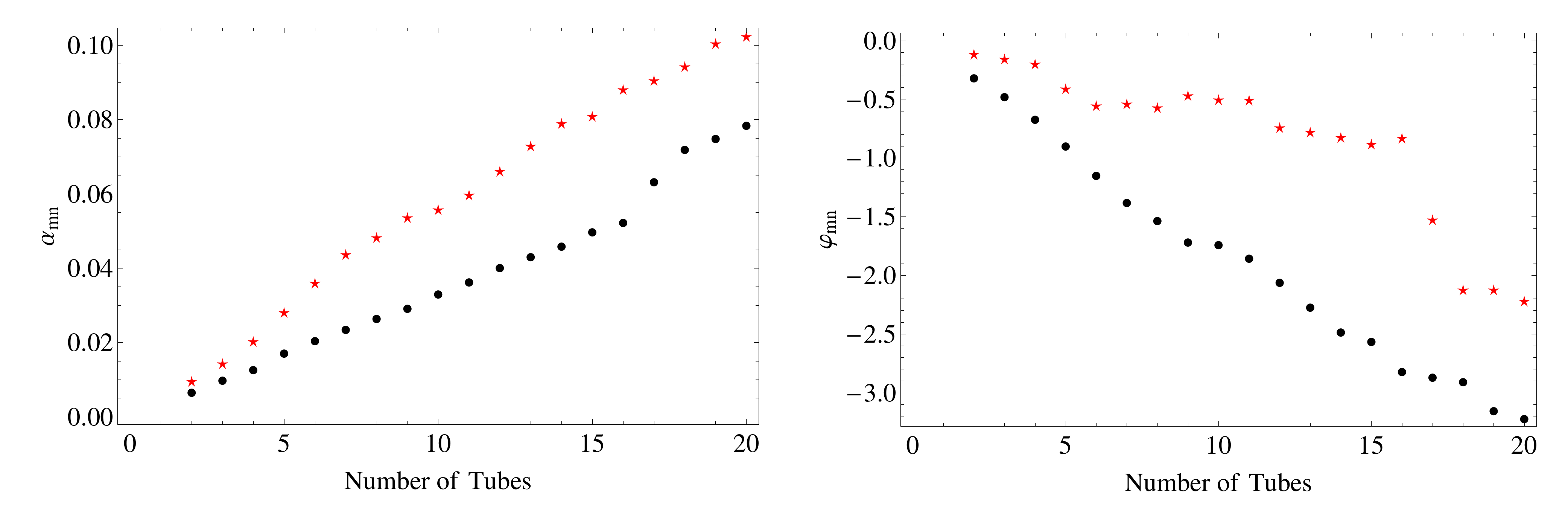}
\caption{The absorption and phase of a $3$~mHz $f$ mode ($m=0$: red star, $m=1$ black dot), as the number of tubes that reside within a circle of radius $2.5$~Mm increases. Each additional tube increases the absorption and negative phase shift. Generally following a linear relationship. We see in this distribution that absorption of the $m=0$ is greater, while the $m=1$ wave has a larger negative phase shift. }
\label{fig:tubenumbers}
\end{figure*}

We conclude this section by exploring how the scattering coefficient changes with an increasing number of flux tubes within 2.5~Mm of the origin. The tubes are identical, randomly distributed and each additional tube is added to the ensemble without moving the others. 
Figure~\ref{fig:tubenumbers} shows the absorption and phase shift for increasing tube numbers for both incident $m=0$ and $m=1$.  The addition of each tube generally increases the absorption, as well as the negative phase shift. Although each added tube is randomly placed, the absorption varies roughly linearly with tube number, and similarly with the phase shift for $m=1$. However, the phase shift of the $m=0$ wave seems to depend quite sensitively on tube position, resulting in a more irregular dependence on tube number.

\section{Discussion and conclusions}\label{sec:concl}
We have extended the work of \citet{hanson_cally_2014} to ensembles of more than two tubes. This study represents a significant step in understanding the seismic behavior of complex ensembles of small magnetic flux tubes threading the solar surface. We have examined cases of both regular symmetric and random flux tube distributions within a stratified atmosphere, allowing scattering between both $m$ and $p_n$ modes. The incoming and outgoing waves are defined in terms of waves centered upon the origin, in a similar fashion to observational Hankel analysis. The caveats to our model are outlined in \citet{hanson_cally_2014}, and while in this paper we address the indirect scattering affects of incident $|m|>1$ waves, the tubes are still assumed to be thin and unable to scatter  pure fluting modes (as defined in their local coordinate system). As such, the direct absorption of fluting modes is still not addressed, due to mathematical complexity. Studies of thicker tubes that can directly interact with these modes may require direct numerical simulation. 

Clearly the thin tube approximation cannot be continued to arbitrary height, as the expansion of the flux tubes eventually sees them thicken to the stage where the assumption breaks down \citep{bogdan_etal_1996}. Our stress-free top boundary condition is a simplistic attempt to avoid this inconvenience, and ignores potential upward losses, both in the flux tubes \citep[though see][]{crouch_cally_1999} and the field-free atmosphere. Upward losses in the field-free region may not be important for low frequencies, where the $p$-modes reflect quite low, but above the acoustic cut-off they do in reality penetrate into the atmosphere.

In this study the definition of absorption is positive by its very nature, as there cannot be more energy emitted than was sent in. \citet{felipe_2013}, as well as the observations of \citet{braun_1995} calculated negative absorption coefficients (emission) in some cases, that were due to the coordinate definition, rather than real emission. We note that this study sends in a pure cylindrical wave, and hence will not produce negative absorption in $m$. In numerical studies (as well as observations), plane waves consist of all $m$ modes and hence interactions between various $m$ in the multiple scattering regime may lead to the perceived `negative' absorption in Hankel analysis.

Having stated these clarifications, let us consider the implications of the symmetric ensemble studies. In these studies the tubes are identical, and positioned in a symmetrical nature around the origin. The three-tube cases are similar in fashion to the numerical studies of \citet{felipe_2013}, while the seven-tube cases are similar to \cite{daiffallah_2014}. The symmetry of the system maintains a mirror distribution of absorption and phase shift between $\pm m$. Complementary to \citet{felipe_2013}, the absorption peaks for the $|m|\le 1$ modes, with $|m|=1$ showing the greatest absorption, but these coefficients diminish rapidly with increasing $|m|$ thereafter. The multiple scattering between the tubes alters the coefficients, primarily for $|m|\le 1$ modes. In this model these results are reasonable, given that the only scattering from a tube is in $|m|\le 1$ and outgoing wave components of $|m|>1$ is purely a fraction of the incoming $|m|\le 1$ wave. As in the study of \citet{hanson_cally_2014}, the near-field enhances the multiple scattering between tubes, leading to greater changes in absorption and phase when compared to the case of far-field interactions alone. Expanding to larger ensembles of six or seven tubes demonstrates similar behavior, with the larger tube numbers altering the absorption and phase in significant ways. Firstly and most clearly, the scattering effects are generally increased as a result of the additional contribution from each tube to the scattered wave field. This is reasonable given the results shown in Figure~\ref{fig:tubenumbers}. Secondly, the presence of additional tubes that are within the near-field of each other creates a significant increase (an order of magnitude) in the change of absorption and phase when compared to the smaller three tube values. Lastly, the addition of the seventh tube reduces $\Delta\alpha$ for the kink modes, while enhancing the negative phase shift of the sausage mode. These three results demonstrate that the addition of more tubes, within close proximity, can change the observable distribution of absorption and phase across $m$.

Interestingly, in symmetric ensembles an incoming wave of even $m$ is unable to generate an outgoing odd $m$ wave, and vice versa. The odd $m$ components scattered by each tube interferes at the origin with the same components generated by a tube located on the other side of the origin. The interference results in no outgoing wave of odd $m$ from the origin, given an incident even $m$ wave. Thus, similar to a single tube that can only scatter the incident $m$ wave, a large symmetric ensemble is unable to scatter waves of odd (or even) $m$ order given the incident wave of even (or odd) order. While truly symmetrical ensembles could not realistically exist, this is an interesting result in the case of near-symmetrical ensembles; we may expect diminished outgoing odd $m$ components from an ingoing even $m$ for roughly symmetric ensembles. In regards to $p_n$ mode scattering, the results are consistent with the observations of \citet{zhao_chou_2013} that the outgoing energy of a $p_{n'}$ mode rapidly diminishes with increasing $n$. The strongest coupling is between the $f$-modes.

We have also investigated random ensembles of tubes. The random positioning of identical tubes breaks symmetry in $m$ of the absorption. The resulting asymmetry is enhanced for higher frequency waves, so they may be more sensitive observational probes. However, phase shows an apparent mirror symmetry across $m$ regardless of tube arrangement. We note that the ensembles show a difference in phase between $\pm m$, but these differences are small compared to the actual values. This apparent symmetry could be attributed to the fact that as more thin tubes are placed randomly within close proximity, the system will approach a closely symmetrical system. In fact, when the system is of a small number of tubes \citep{hanson_cally_2014}, or contains thicker tubes \citep{felipe_2013}, the asymmetric nature is more apparent in the phase data. 

Given these results and taking into account the simplicity of the model, how can we use this model to interpret observable parameters?  \citet{braun_1995} was amongst the first to determine the absorption and phase of cylindrical waves within magnetic plage regions, finding significant measurable absorption, while \citet{braun_birch_2008} found that plage produces considerably less shift than penumbrae and umbrae. We have shown here that within closely packed ensembles the multiple scattering regime enhances both absorption and phase shift. In fact the larger the ensemble, or higher the frequency, the greater the absorption. This is a reasonable result, given that each additional tube will absorb more of the incident wave, and will have this absorption coefficient enhanced by the scattered wave field from nearby tubes.  Within large enough ensembles the absorption coefficients should be measurable above any noise. The spatial distribution of tubes appears to have a greater effect on absorption at higher frequencies (5~mHz), which therefore may be more useful observational probes.
Also the absorption profiles are heavily altered by the characteristics of the individual tubes (in this study $\beta$), and may be used to indicate the internal constitution. Our obtained phase shifts are small, possibly too small for measurable certainty (especially at 3~mHz), and thus are not in contradiction to observation. Given observations with an appropriate spatial/temporal resolution, the outgoing phase may be used for probing sub-surface structures, as we have shown that individual tube characteristics as well as the number of tubes will affect the outgoing phase. This model is a significant step in an analytical approach to the multiple scattering regime. As the aim of scattering studies is to constrain sub-surface structures with scattered wave field data, both numerical and analytical models must continue to be improved to better interpret observational results.

\bigskip
\bibliographystyle{apj}        
\bibliography{References}
\end{document}